\documentclass[aps,prl]{revtex4}
\newcommand{\beq}{\begin{equation}} \newcommand{\eeq}{\end{equation}}
\newcommand{\bea}{\begin{eqnarray}} \newcommand{\eea}{\end{eqnarray}}
\newcommand{\bear}{\begin{eqnarray*}} \newcommand{\eear}{\end{eqnarray*}}
\newcommand{\lb}{\label} 
\newcommand{\rf}[1]{(\ref{#1})}   
\usepackage[pdftex]{graphicx}
\usepackage{epstopdf}
\usepackage{amsmath}
\usepackage{amsfonts}
\usepackage{amssymb}
\usepackage{makeidx}

\begin{document}

\title {Integrable inhomogeneous spin chains in generalized Lunin-Maldacena backgrounds}

\author{Matheus Jatkoske Lazo\footnote{lazo@smail.ufsm.br}} 
\address{Centro Tecnol\'ogico de Alegrete, UFSM/Unipampa, \\ Alegrete, RS, Brazil. \\ Programa de P\'os-gradua\c c\~ao em F\'\i sica, UFSM, 97111-900, \\ Santa Maria, RS, Brazil}

\begin{abstract}

We obtain through a Matrix Product Ansatz the exact solution of the most
general inhomogeneous spin chain with nearest neighbor interaction and with
$U(1)^2$ and $U(1)^3$ symmetries. These models are related to the one loop
mixing matrix of the Leigh-Strassler deformed $N=4$ SYM theory, dual to type
IIB string theory in the generalized Lunin-Maldacena backgrounds, in the
sectors of two and three kinds of fields, respectively. The solutions
presented here generalizes the results obtained by the author in a previous
work for homogeneous spins chains with $U(1)^N$ symmetries in the sectors of
$N=2$ and $N=3$.



\end{abstract}

\maketitle

\noindent

\section{Introduction}

String theory was first introduced in the 60's years, as a tentative to describe
the big amount of mesons and hadrons discovered in particles accelerators. In
this theory the particles are seen as different oscillations modes of the
strings and it was able to describe some spectral characteristics of
hadrons. Although this success, the original string theory is not able to
explain many physical phenomenon mediated by strong interactions and nowadays
the most successful theory to describe particle physics is the Quantum 
Cromodynamics (QCD). On the other hand, despite string theory has reemerged in the
last three decades as a promising candidate for a quantum theory of all
known interactions, its validity to describe nature has generated heated
discussions since the theory is untestable due to the experimental
impossibility to reach the tremendous energies found at the Plank scale. 
So, what is the actual relevance of string theory for real world physics?
Either string theory is the correct theory to describe the world or not,
remarkable resulties showed that string theory can be seen as a different
formalism to quantum field theory. In QCD, it is not possible to obtain a
satisfactory quantitative description in the small energy regime, when the 
coupling constant is very strong. In this regime, numerical calculation on the 
lattice is the best tool for study physical models. On the other hand, it was
observed by 't Hooft \cite{Hooft} that the theory is simplified when the number N 
of colors is very high. 't Hooft derived a relationship between the
topological structure of a Feynman graph and its $N$ dependence. When $1/N$ is 
interpreted as a coupling constant, an expansion in $1/N$ is similar to an 
expansion in a generic interacting string theory, resulting in a relation
between strings and planar diagrams. If the $N=3$ can be regarded as a large $N$,
it explain why the string models of 60's years was able to give the right
relation for some spectral characteristics of hadrons. More recently a
remarkable result arouses new interest in the duality between string theory and
quantum field theory. Maldacena conjectured that IIB string theory on
the curved background $AdS_5 \times S^5$ (anti-de Sitter and sphere spaces) 
should be equivalent to $N = 4$ Super Yang-Mills (SYM)
\cite{Maldacena,GKP,Witten}. This conjecture, AdS/CFT, relates operators, 
states, correlation functions and dynamics of both theories. One of the most 
important results of this conjecture predicts that the spectrum of scaling 
dimension operator of gauge invariant operators, in the conformal field theory,
should coincide with the spectrum of energies $E$ of string states.
Furthermore, this correspondence relates the weak coupling constant regime, 
in the gauge theory, with the strong coupling constant ones, in the string 
theory. 

The Maldacena' s conjecture need yet to be fully proved and since the
discovery of the relation between the planar dilatation operator of the 
${\cal{N}}=4$ SYM with an integrable $so(6)$ quantum spin chains \cite{MZ}, 
integrability has played a prominent role in the exploration of the
Maldacena's  correspondence. The study of the planar dilatation operator's 
integrability is very important because it not only enable us to test the 
Maldacena's correspondence as it is an generator of nontrivial integrable 
models. Exactly solvable models are of interest in high energy physics,
condensed matter physics, statistical mechanics and mathematics since 
the pioneering work of Hans Bethe \cite{bethe} (see, e.g., 
\cite{baxter,revkore,revessler,revschlo} for reviews). According to 
this ansatz the amplitudes of the eigenfunction are expressed by a 
nonlinear combination of properly defined plane waves. On the other hand, 
in the last two decades several different ansatz were introduced in the 
literature under the general name of matrix product ansatz (MPA). The first 
formulation was done for the description of the ground-state eigenfunction 
of some special non-integrable quantum chains, the so called valence-bond 
solid models \cite{affleck,arovas,fannes,klumper}. The MPA becomes also a 
successful tool for the exact calculation of the stationary probability 
distribution of some stochastic one dimensional systems 
\cite{derr2,derr1,alcritda}. An extension of this last MPA, called dynamical 
MPA was introduced in \cite{schutz1,schutz2} and extended in \cite{popkov1}. 
This last ansatz gives the time-dependent probability distribution for some 
exact integrable systems. The MPA we are going to use in this paper was 
introduced in \cite{alclazo1,alclazo2,alclazo3,alclazo4}. This ansatz was 
applied with success in the evaluation of the spectra of several integrable 
quantum Hamiltonians \cite{alclazo1,alclazo2,alclazo3}, transfer matrices 
\cite{ice,lazo,alclazo5} and the time-evolution operator of stochastic systems 
\cite{alclazo4}. According to this ansatz, the amplitudes of the
eigenfunctions are given in terms of a product of matrices where the matrices 
obey appropriated algebraic relations. In the case of the Bethe ansatz the 
spectral parameters and the amplitudes of the plane waves are fixed, apart 
from a normalization constant, by the eigenvalue equation of the Hamiltonian 
or transfer matrix. On the other hand, in the MPA the eigenvalue equation 
fixes the commutation relations of the matrices defining the ansatz. In such 
case the spectrum of the Hamiltonian or transfer matrix, and the corresponding 
eigenfunctions, can be computed in a purely algebraic way.

In the present paper we obtain through a MPA the exact solution of the most 
general inhomogeneous spin chain with nearest neighbor interaction and with 
$U(1)^2$ and $U(1)^3$ symmetries. This model is related to the one loop
dilatation operator in deformed Lunin-Maldacena backgrounds
\cite{LuninMaldacena} and conformal field theories with deformations 
\cite{LeighStrassler}. The solutions presented here generalizes 
the results obtained for homogeneous spins chains with $U(1)^N$ symmetries in
the sectors of $N=2$ and $N=3$ \cite{lazo2,FT}. In this model the coupling
interaction between neighbor sites are not a constant as it is in the
homogeneous model studied in \cite{lazo2,FT}.


\section{The inhomogeneous spin chains}

We consider here the most general inhomogeneous $3$-state spin chain with
nearest neighbor interaction, periodic boundary condition and $U(1)^3$ 
symmetry. The $U(1)^3$ symmetry imply that the Hamiltonian describing the 
time evolution of this spin chain conserves the number of states of each 
type. By denoting the basis of states at a given site as $|\alpha\rangle$ 
($\alpha=1,2,3$), the Hamiltonian in a periodic lattice with $L$ sites 
takes the form 
\beq
H=\sum_{j=1}^L \left(
\sum_{\alpha \ne \beta =1}^3
\Gamma_{\beta\;\alpha}^{\alpha\;\beta}(j,j+1)E_j^{\beta\;\alpha}E_{j+1}^{\alpha\;\beta}+\sum_{\alpha
  ,\beta=1}^3 \Gamma_{\alpha\;\beta}^{\alpha\;\beta}(j,j+1)E_j^{\alpha\;\alpha}E_{j+1}^{\beta\;\beta}\right),
\lb{3}
\eeq
where $E^{\alpha\;\beta}$ are $3 \times 3$ Wyel matrices with elements 
$\left(E^{\alpha\;\beta}\right)_{i,j}=\delta_{\alpha,i}\delta_{\beta,j}$
($\alpha,\beta=1,2,3$). 
While the first term in the right hand side of \rf{3} acts over neighbor
sites exchanging its configuration $|\alpha\rangle_j \otimes|\beta\rangle_{j+1}
\rightarrow |\beta\rangle_j \otimes |\alpha\rangle_{j+1}$ with rate
$\Gamma_{\beta\;\alpha}^{\alpha\;\beta}(j,j+1)$, the second one is a diagonal
operator with weight $\Gamma_{\alpha\;\beta}^{\alpha\;\beta}(j,j+1)$. The
Hamiltonian \rf{3} in a particular case
($\Gamma_{\gamma\;\delta}^{\alpha\;\beta}(j,j+1)=\Gamma_{\gamma\;\delta}^{\alpha\;\beta}$
constant) contain the homogeneous integrable spin-$1$ chain, related to the one loop
dilatation operator in deformed Lunin-Maldacena backgrounds, studied by Freyhult, Kristjansen and Mansson \cite{FT} and generalized in \cite{lazo2}. The eigenfunctions for
\rf{3} can be construct as 
\beq
\lb{4}
|\Psi_L\rangle=\sum_{\alpha_1,...,\alpha_L}^* \Psi^{\alpha_1...\alpha_L}|\alpha_1,...,\alpha_L\rangle
\;\;\; (\alpha_j=1,2,3),
\eeq
where the symbol $(*)$ in the sum denotes the restriction to the sets
$\{\alpha_1,...,\alpha_L\}$ with the same number $n_{\alpha}$ of spins in
configuration $\alpha$ and $\Psi^{\alpha_1...\alpha_L}$ is a constant.


\section{The MPA}

In order to formulate a MPA for the Hamiltonian \rf{3}, we make a one-to-one
correspondence between configurations of spins and products of abstract
matrices. This matrix product is construct by making a correspondence between
a site $j$ in the chain with spin configuration $\alpha=1,2,3$ and a matrix
$A_j^{(\alpha)}$. Our MPA asserts that the components of the amplitude of the
eigenfunction $\Psi^{\alpha_1...\alpha_L}$ in \rf{4} are obtained by
associating them to a products of these matrices $A_j^{(\alpha)}$. Actually
$A_j^{(\alpha)}$ are abstract operators with an associative product. A well
defined eigenfunction is obtained, apart from a normalization factor, if all
the amplitudes are related uniquely. Equivalently, in the subset of words
(products of matrices) in the algebra containing $n_{\alpha}$
($n_1+n_2+n_3=L$) matrices $A_j^{(\alpha)}$ there exists only a single
independent word. The relation between any two words gives the ratio between
the corresponding amplitudes of the components of the eigenfunction
$|\Psi_L\rangle$. To formulated the ansatz we can choose any uniform operation
on the matrix products that gives a non-zero scalar to make a map between the
amplitudes $\Psi^{\alpha_1...\alpha_L}$ in \rf{4} and a matrix product
\footnote{Actually this uniform operation need to satisfy the symmetries of
  the model. In \rf{lazo2} we choose for this map the trace operation since
  the model were homogeneous and defined in a periodic chain. However in the
  present paper due to the inhomogeneities the trace operator do not work
  although we introduce another auxiliary matrix as in \rf{alclazo1,alclazo2}.}:
\beq
\lb{5}
\Psi^{\alpha_1...\alpha_L}\Longleftrightarrow A_1^{(\alpha_1)}A_2^{(\alpha_2)}\cdots A_L^{(\alpha_L)} \;\;\; (\alpha_j=1,2,3).
\eeq

It is obvious that the $3$ states $|\alpha...\alpha\rangle$ ($\alpha=1,2,3$)
are all eigenstates of the Hamiltonian \rf{3}. In the following we shall
choose $|1...1\rangle$ as our reference state. The Hamiltonian \rf{3} when
applied to the components of the eigenfunction \rf{4} where we do not have
spins configurations $|\alpha\rangle$ ($\alpha=2,3$) at nearest neighbor
sites and at boundary positions give us the constraints, for the amplitudes
$\Psi^{\alpha_1...\alpha_L}$
\rf{5},
\bea
\label{6}
&&\!\!\!\!\!\!\!\varepsilon_n A^{x_1-1}A_{x_1}^{(\alpha_1)} A^{x_2-x_1-1} \cdots
A_{x_{j-1}}^{(\alpha_{j-1})} A^{x_j-x_{j-1}-1}A_{x_j}^{(\alpha_j)} A^{x_{j+1}-x_j-1}A_{x_{j+1}}^{(\alpha_{j+1})}\cdots A_{x_n}^{(\alpha_n)} A^{L-x_n}\nonumber \\
&& \;\;\;\;\;\; =\sum_{j=1}^n \left[ \Gamma_{1 \;\; \alpha_j}^{\alpha_j \;1}(j,j+1) A^{x_1-1}A_{x_1}^{(\alpha_1)}A^{x_2-x_1-1} \cdots
A_{x_{j-1}}^{(\alpha_{j-1})}A^{x_j-x_{j-1}-2}A_{x_j-1}^{(\alpha_j)}A^{x_{j+1}-x_j}A_{x_{j+1}}^{(\alpha_{j+1})}\cdots A_{x_n}^{(\alpha_n)}A^{L-x_n}\right.\nonumber \\
&& \;\;\;\;\;\; +\left. \Gamma_{\alpha_{j+1} \; 1}^{1\;\; \alpha_{j+1}}(j,j+1)  A^{x_1-1}A_{x_1}^{(\alpha_1)}A^{x_2-x_1-1} \cdots A_{x_{j-1}}^{(\alpha_{j-1})}A^{x_j-x_{j-1}}A_{x_j+1}^{(\alpha_j)}A^{x_{j+1}-x_j-2}A_{x_{j+1}}^{(\alpha_{j+1})}\cdots
A_{x_n}^{(\alpha_n)}A^{L-x_n} \right]\nonumber \\
&&\;\;\;\;\;\;+\sum_{l=1}^L \left[\Gamma_{\alpha_l \; 1}^{\alpha_l \; 1}(l,l+1)+ \Gamma_{1\; \alpha_{l+1}}^{1\;
    \alpha_{l+1}}(l,l+1)\right] A^{x_1-1}A_{x_1}^{(\alpha_1)} A^{x_2-x_1-1} \cdots
A_{x_n}^{(\alpha_n)} A^{L-x_n} \;\;\;\;\; (\alpha_j=1,2,3),
\eea
where $\varepsilon_n$ is the energy of the eigenfunction \rf{4}, $A\equiv
A_x^{(1)}$, $n=n_2+n_3$, and $x_1,...,x_n$ are the position in the spin chain
where we have a state configuration $|\alpha\neq 1\rangle$. A convenient
solution of this last equation is obtained by identifying the matrices
$A_x^{(\alpha)}$ ($\alpha=2,3$) as composed by spectral-parameter-dependent
matrices. The distinguibility of states configurations allows two types of 
solutions. The standard solution is obtained if each of the matrices
$A_x^{(\alpha)}$ ($\alpha=2,3$) is composed of $n=n_2+n_3$ spectral parameter
dependent matrices \cite{alclazo2,alclazo4,lazo2}. A second class of solutions
is obtained if matrices $A_x^{(\alpha)}$ ($\alpha=2,3$) with different $\alpha$
value are composed of by distinct sets of spectral parameters matrices \cite{alclazo2,
  lazo2}. Here we will consider only the standard solution but our model
can be easily extended to the second class problem. In the present case, the 
matrices $A_x^{(\alpha)}$ ($\alpha=2,3$) can be written in terms of the matrix
$A$ and $n=n_2+n_3$ spectral parameter dependent matrices
$A^{(\alpha)}_{x,k_j}$ \footnote{The most general relation
  $A_x^{(\alpha)}=\sum_{j=1}^n A^{a} A^{(\alpha)}_{x,k_j}A^{b}$
  could be used. However \rf{6} is more convenient since otherwise the
  $S$-matrix in \rf{12} will depend on $a$ an $b$}:
\beq
\lb{7}
A_x^{(\alpha)}=\sum_{j=1}^n A^{(\alpha)}_{x,k_j}A, \;\;\;\; (\alpha=2,3),
\eeq
where the matrices $A^{(\alpha)}_{x,k_j}$ satisfy the following commutation
relations with the matrix $A$:
\beq
\lb{8}
A^{(\alpha)}_{x,k_j}A=g_\alpha(x,x+1)e^{ik_j}AA^{(\alpha)}_{x+1,k_j},  \;\;\; (\alpha=2,...,N),  \;\;\; (j=1,...,n),
\eeq
the parameters $k_j$ ($j=1,...,n$) are in general complex numbers unknown {\it
  a priori}, and $g_\alpha(x,x+1)$ is a constant. The energy $\varepsilon_n$
is obtained by inserting \rf{7} in \rf{6}, by using \rf{8} and imposing that
$\varepsilon_n$ is a symmetric function on the spectral parameters
\footnote{There is another possibility to symmetrize the eigenvalue. It is to
  make all diagonal coupling equal
  $\Gamma_{\alpha\;\beta}^{\alpha\;\beta}(x,x+1)=\Gamma_{1\;1}^{1\;1}(x,x+1)$
  ($\alpha, \beta =1,2,3$). In this case the diagonal interaction can be
  eliminated by adding a constant in the Hamiltonian \rf{3} and the problem
  reduces to a free fermion model.}
\beq
\lb{9}
\varepsilon_n = \sum_{j=1}^n \left(\Gamma_{1\;2}^{2\;1}e^{ik_j}+ \Gamma_{2\;1}^{1\;2}e^{-ik_j}\right) + 
\sum_{\alpha=2}^3 n_{\alpha}\left(\Gamma_{1\;\alpha}^{1\;\alpha}+\Gamma_{\alpha\;1}^{\alpha\;1}\right)+(L-2n)\Gamma_{1\;1}^{1\;1},
\eeq
where we need to impose
\beq
g_\alpha(x,x+1)=\frac{\Gamma_{1\;2}^{2\;1}}{\Gamma_{1\;\alpha}^{\alpha\;1}(x,x+1)}=\frac{\Gamma_{\alpha\;1}^{1\;\alpha}(x,x+1)}{\Gamma_{2\;1}^{1\;2}}
\;\;\; (\alpha=2,3)
\lb{10a}
\eeq
and
\beq
\Gamma_{1\;1}^{1\;1}(x,x+1)=\Gamma_{1\;1}^{1\;1},\;\;\;\;\;
\Gamma_{1\;\alpha}^{1\;\alpha}(x,x+1)=\Gamma_{1\;\alpha}^{1\;\alpha},\;\;\;\;\;
\Gamma_{\alpha\;1}^{\alpha\;1}(x,x+1)=\Gamma_{\alpha\;1}^{\alpha\;1} \;\;\; (\alpha=2,3),
\lb{10b}
\eeq
where $\Gamma_{1\;2}^{2\;1}$, $\Gamma_{2\;1}^{1\;2}$, $\Gamma_{1\;1}^{1\;1}$,
$\Gamma_{1\;\alpha}^{1\;\alpha}$ and $\Gamma_{\alpha\;1}^{\alpha\;1}$ are constants.

The relations coming from the eigenvalue equation for configurations where we
have two spins configurations $|\alpha\rangle$ ($\alpha=2,...,N$) at nearest
neighbor sites and are not located at boundary positions given us 
\bea
\lb{11}
&&\sum_{j,l=2}^n\left[\Gamma_{2\;1}^{1\;2}+\Gamma_{1\;2}^{2\;1}e^{i(k_j+k_l)}+(\Gamma_{\alpha\;1}^{\alpha\;1}+\Gamma_{1\;\alpha}^{1\;\alpha}-\Gamma_{1\;1}^{1\;1}-\Gamma_{\alpha\;\alpha}^{\alpha\;\alpha}(x,x+1))e^{ik_j}
\right]A^{(\alpha)}_{y,k_j}A^{(\alpha)}_{y,k_l}=0, \nonumber \\
&&\sum_{j,l=2}^n\left[\Gamma_{2\;1}^{1\;2}+\Gamma_{1\;2}^{2\;1}e^{i(k_j+k_l)}+(\Gamma_{\alpha\;1}^{\alpha\;1}+\Gamma_{1\;\beta}^{1\;\beta}-\Gamma_{1\;1}^{1\;1}-\Gamma_{\alpha\;\beta}^{\alpha\;\beta}(x,x+1))e^{ik_j}
\right]A^{(\alpha)}_{y,k_j}A^{(\beta)}_{y,k_l}=  \\
&& \;\;\;\;\;\;\;\;\;\;\;\;\;\;\;\;\;\;\;\;\;\;\;\;\;\;\;\;\;\;\;\;\;\;\;\;\;\;\;\;\;\;\;\;\;\;\;\;\;\;\;\;\;\;\;\;\;\;\;\;\;
\sum_{j,l=2}^n\frac{\Gamma_{1\;\alpha}^{\alpha\;1}(x,x+1)}{\Gamma_{1\;\beta}^{\beta\;1}(x,x+1)}\Gamma_{\alpha\;\beta}^{\beta\;\alpha}(x,x+1)e^{ik_l}A^{(\beta)}_{y,k_j}A^{(\alpha)}_{y,k_l}\;\;\;
(\alpha \neq \beta), \nonumber
\eea
where we have used \rf{5}, \rf{7}-\rf{10b} and $y=1,...,L$. The relations
\rf{11} should be satisfied for all $x,y=1,...L$. This is possible if we have
constants $\Gamma_{\alpha\;\alpha}^{\alpha\;\alpha}$,
$\Gamma_{\alpha\;\beta}^{\alpha\;\beta}$,
$\Gamma_{\alpha\;\beta}^{\beta\;\alpha}$ and if
\beq
\Gamma_{\alpha\;\alpha}^{\alpha\;\alpha}(x,x+1)=\Gamma_{\alpha\;\alpha}^{\alpha\;\alpha},
\;\;\;\;\;
\Gamma_{\alpha\;\beta}^{\alpha\;\beta}(x,x+1)=\Gamma_{\alpha\;\beta}^{\alpha\;\beta},
\;\;\;\;\; \Gamma_{\alpha\;\beta}^{\beta\;\alpha}(x,x+1)=\frac{\Gamma_{1\;\beta}^{\beta\;1}(x,x+1)}{\Gamma_{1\;\alpha}^{\alpha\;1}(x,x+1)} \Gamma_{\alpha\;\beta}^{\beta\;\alpha}.
\lb{11b}
\eeq
Finally, the relations \rf{11} fix the algebraic relations among the matrices $A^{(\alpha)}_{x,k_j}$ $(\alpha=2,3)$:
\beq
\lb{12}
A^{(\alpha)}_{x,k_j}A^{(\beta)}_{x,k_l}=\sum_{\alpha',\beta'=2}^3 S^{\alpha \; \beta}_{\beta' \; \alpha'}(k_j,k_l)A^{(\alpha')}_{x,k_l}A^{(\beta' )}_{x,k_j},  \;\;\;A^{(\alpha)}_{x,k_j}A^{(\beta)}_{x,k_j}=0 \;\;\; (l \neq j =1,...,n).
\eeq
Relations \rf{8} and \rf{12} define completely the algebra whose structural
constants are the $S$-matrix of the spin-$\frac{3}{2}$ model
\cite{alclazo1,alclazo2,lazo2}. Since the several components of the
wavefunction should be uniquely related, the above algebra should be
associative. This associativity implies that the above $S$-matrix should
satisfy the Yang-Baxter relations \cite{baxter,Yang2}, which is indeed the
case \cite{lazo2}. The components of the wavefunction corresponding to the
configurations where we have three or four particles in next-neigbouring sites
would give in principle new relations involving three or four matrices
$A^{(\alpha)}_{x,k_j}$. These new relations are however consequences of the
above relations \rf{8} and \rf{12}. It is important to mention that in the
sector $n_2=0$ or $n_3=0$ (symmetry $U(1)^2$) the Hamiltonian \rf{3} reduces
to a inhomogeneous version of the well known asymmetric XXZ model
\cite{yangyang}.

In order to complete our solutions through the MPA \rf{5} we should fix the
spectral parameters, or momenta, $k_1,\ldots, k_n$. Theses parameter are fixed
 from the configurations where we have a spin configuration $\alpha=2,3$ at
 boundary positions ($x_1=1$ or $x_n=L$). By using the algebraic relations
 \rf{7}, \rf{8} and \rf{12} we obtain the relation
\beq
A_{1,k_1}^{(\alpha_1)}\cdots  A_{n,k_n}^{(\alpha_n)}A^L = e^{ik_jL} \sum_{\alpha_1',\ldots,\alpha_n'=2}^N \langle
\alpha_1,\ldots,\alpha_n|{\cal{T}}^{(n)}|\alpha_1',\ldots,\alpha_n'\rangle A_{1,k_1}^{(\alpha_1')}\cdots  \alpha_{n,k_n}^{(\alpha_n')}A^L,
\lb{15}
\eeq
where we have used the identity (see \cite{lazo2})
\beq
\sum_{\alpha_j'',\alpha_{j+1}''}S_{\alpha_j'\;\alpha_j''}^{\alpha_j\;\alpha_{j+1}''}(k_j,k_j)=-1,
\lb{16}
\eeq
and
\beq \langle \alpha_1,\ldots,\alpha_n|{\cal{T}}^{(n)}|\alpha_1',\ldots,\alpha_n'\rangle = \sum_{\alpha_1'',\ldots,\alpha_n''} \left\{ S_{\alpha_1'\;\alpha_1''}^{\alpha_1\;\alpha_2''}(k_1,k_j)\cdots S_{\alpha_j'\;\alpha_j''}^{\alpha_j\;\alpha_{j+1}''}(k_j,k_j)\cdots S_{\alpha_n'\;\alpha_n''}^{\alpha_n\;\alpha_1''}(k_n,k_j)\phi(\alpha_1'') \right\}
\lb{17}
\eeq
where $\phi(\alpha_1'')=\prod_{x=1}^L g_{\alpha_1''}(x,x+1)$, is a $(2)^n
\times (2)^n$-dimensional transfer matrix of an inhomogeneous vertex model
(inhomogeneities $\{k_l \}$) with Boltzmann weights given by the $S$-matrix
elements defined in \rf{12}. The model is defined on a cylinder of perimeter
$n$ with a seam along its axis producing the twisted boundary condition
\beq
\lb{17b}
S_{\alpha_n'\;\alpha_n''}^{\alpha_n\;\alpha_{n+1}''}(k_n,k_j)=S_{\alpha_n'\;\alpha_n''}^{\alpha_n\;\alpha_1''}(k_n,k_j)\phi(\alpha_1'').
\eeq
Finally relation \rf{15} with \rf{17b} give us the constraints for the spectral parameters:
\beq
e^{-ik_jL}=\Lambda^{(n)}(k_j,\{k_l \}) \;\;\; (j=1,\ldots,n), 
\lb{18}
\eeq
where $\Lambda^{(n)}(k_j,\{k_l \})$ are the eigenvalues of the transfer matrix \rf{17}. The condition \rf{18} leads to the problem of evaluation the eigenvalues of the inhomogeneous transfer matrix \rf{17}. This can be done through the algebraic Bethe  ansatz \cite{kulish} or the coordinate Bethe
ansatz (see \cite{alcbar4} and \cite{bjp4} for example).

\section{Discussion and conclusions}

We solve through a MPA the most general inhomogeneous $3$-state spin chain
with $U(1)^3$ symmetry and nearest neighbor interaction. We found that the
coupling constants in \rf{3} should satisfy the constraints \rf{10a}, \rf{10b}
and \rf{11b} in order to make the Hamiltonian \rf{3} integrable. It is important to
mention that different from the homogeneous spin chain, where the eigenstates
\rf{4} are also eigenstates of the translation operator due to the periodic
boundary condition, in the inhomogeneous model the eigenstates do not have a
defined momentum. It happens due to the inhomogeneities that broke the
translational invariance of the system. The study of this new model can
be of interest in the context of both AdS/CFT and condensed matter physics
since it is related to the one loop dilatation operator in deformed
Lunin-Maldacena backgrounds \cite{LuninMaldacena}, conformal field theories
with deformations \cite{LeighStrassler} and inhomogeneous spin chains. Another
quite interesting problem for the future concerns the formulation of the MPA
for the case where we have open boundary conditions, as well as for quantum
chains with no global conservation law such as the XYZ model, the 8-vertex
model or the case where the quantum chains are defined on open lattices with
non-diagonal boundary fields.

\section{Acknowledgements}

This work has been supported by CAPES (Brazilian agencies).

\end{document}